\def\ptitle{Soft-core Coulomb potentials and Heun's differential equation}
\def\half{\frac{1}{2}}

\documentclass{revtex4}
\usepackage{amsmath}
\usepackage{graphicx}
\usepackage{rotating}
\begin{document}


\title{\ptitle}

\author{Richard L. Hall$^1$, Nasser Saad$^2$, and K. D. Sen$^3$}

\address{$^1$ Department of Mathematics and Statistics, Concordia University,
1455 de Maisonneuve Boulevard West, Montr\'eal, Qu\'ebec, Canada
H3G 1M8}
\address{$^2$ Department of Mathematics and Statistics,
University of Prince Edward Island, 550 University Avenue,
Charlottetown, PEI, Canada C1A 4P3.}
\address{$^3$ School of Chemistry, University  of Hyderabad 500046, India.}

\email{rhall@mathstat.concordia.ca} \email{nsaad@upei.ca}
\email{sensc@uohyd.ernet.in}
\medskip
\begin{abstract}
\noindent Schr\"odinger's equation with the attractive potential $V(r) = -{Z}/{(r^q+\beta^q)^{\frac{1}{q}}}$ , $Z>0,~\beta>0,~q \ge 1$,
is shown, for general values of the parameters $Z$ and $\beta$, to be reducible to the confluent Heun equation in the case $q=1$,
and to the generalized Heun equation in case $q=2$. In a formulation with correct asymptotics, the eigenstates are
specified {\it a priori} up to an unknown factor. In certain special cases this factor becomes a polynomial.
The Asymptotic Iteration Method is used either to find the polynomial factor and the associated eigenvalue
explicitly, or to construct accurate approximations for them. Detail solutions for both cases are provided.

 \end{abstract}
 \maketitle



\maketitle
\section{Introduction}
\noindent Analytic wave functions corresponding to solutions of
Schr\"odinger's time-independent equation $H\Psi = E\Psi$ for standard
quantum-mechanical systems, such as the hydrogen atom and the harmonic
oscillator, have usually been constructed by transforming the
original equations in to a differential equation whose solutions
are known in terms of certain special functions. Indeed the
general solutions of the resulting Kummer equation corresponding
to such model systems, as expressed in the form of confluent
hypergeometric functions, have led to a deeper understanding of
these systems, under both free \cite{BETH,MOSH} and spatially confined
conditions \cite{SPM}. In this paper we consider the model quantum
system defined by the Hamiltonian (in atomic units $m = \hbar = e
= 1$)
\begin{equation}\label{sec1eq1}
H = -\half\Delta + V_q(r),\quad V_q(r) =
-\frac{Z}{(r^q+\beta^q)^{\frac{1}{q}}}.
\end{equation}
The potential $V_q(r)$ represents a family of soft-core (truncated) Coulomb potentials,
 which are useful as model potentials in atomic and molecular physics. The bound states are
obtained in terms of three potential parameters: the coupling
$Z>0,$ the cut-off parameter $\beta >0,$ and the power parameter
$q \ge 1.$  The cases $q=1$ and $2$ are of special physical
significance \cite{MP,P1,SVD,DMVD,SR,FF0,CM,OM,MO}.  The potential
$V_1$ represents the potential due to a smeared charge and is
useful in describing mesonic atoms. The potential $V_2$ is similar
to the shape of the potential due to a finite nucleus and
experienced by the muon in a muonic atom. Extensive applications
of the soft-core Coulomb potential, $V_2$, have been made through
model calculations corresponding to the interaction of intense
laser fields with atoms \cite{LM,JHE1,JHE3,PLK,CWC,CK}. The
parameter $\beta$ can be related to the strength of the laser
field, with the range $\beta=20-40$ covering the experimental
laser field strengths \cite{LM}.\medskip

In an earlier paper \cite{HSSC09}, we have carried out a general analysis of
the characteristic features of the energies and wave functions of
the complete family of soft Coulomb potentials defined by $V_q$. 
The Schr\"odinger operator $H$ is bounded below. This may be shown immediately
by an application of the operator inequality \cite{RS2,GS}
$-\Delta > 1/(4r^2)$, which yields the general spectral bound
\[
E >\min_{r > 0}\left[\frac{P^2}{2r^2} + V_q(r)\right],\quad P = \half.
\]
We showed in Ref. \cite{HSSC09} that
general upper and lower estimates for
all the discrete eigenvalues can be expressed in this form (for suitable $P$),
 with the aid of envelope theory \cite{env1,env2,env3,env4}.
If the exact eigenvalues of $H$ are written $E(Z,\beta,q)$, then these spectral functions are
 monotone in each parameter, decreasing in $Z$ and $q$,
and increasing in $\beta$. Thus
\[
\frac{\partial E}{\partial Z} < 0,\quad \frac{\partial E}{\partial
\beta} > 0,\quad {\rm and}\quad \frac{\partial E}{\partial q} < 0.
\]
Moreover, the following general scaling law is obeyed:
\begin{equation}\label{sec1eq2}
E(Z,\beta,q) = Z^2E(1,Z\beta,q) = \frac{1}{\beta^2}E(Z\beta,1,q).
\end{equation}
In the limit as $q\rightarrow \infty$, the potential descends to the cut-off Coulomb potential $V_{\infty}$ given by
\[
\lim_{q\rightarrow\infty} V(r) = V_{\infty}(r) = \left\{
\begin{array}{l l}
-\frac{Z}{\beta},  &\text{if $r < \beta$;}\\
\\
-\frac{Z}{r},      &\text{if $r \ge \beta$.}
\end{array}
\right.
\]
This much is known for the whole class of problems.\medskip

Mehta and Patil \cite{MP} have presented
analytical solutions for the $s$-state eigenvalues corresponding
to the $V_1$ potential. Patil \cite{P1} has also discussed the
analyticity of the scattering phase shifts for two particles
interacting through the potentials
 $V_q$ with $q=1$ and $q=2$. Singh {\it et al} \cite{SVD} have reported
 a large number of eigenvalues for the states $1s$ to $4f$ corresponding to
 $V_1$ and $V_2$ for a fixed value of $Z$; these values were obtained by
 the numerical solution of Eq.(1) for $Z = 1.$ The scaling law (\ref{sec1eq2})
 extends their application to other values of $Z.$
  Exact bound-state solutions of $V_1$
 have been considered earlier \cite{DMVD,SR,FF0,CM}, in which only a
 limited number of states with a specific choice of ${\ell}=0\dots 3$ have been treated.
 To our knowledge, no such study for $q \geq 2 $ has been
 reported so far.\medskip

The principal results of the present  paper are as follows. We
show that Schr\"odinger's equation for the two physically
important cases, $q=1$ and $q = 2$, can both be solved
analytically, in terms of the Heun differential equation and the
generalized Heun equation, respectively. Interestingly, in a
suitable formulation, we show that exact solutions exist in which
a factor in the wave function becomes a polynomial. It is shown
that such solutions, and the corresponding exact eigenvalues, can
be found explicitly by the use of an iterative method called the
Asymptotic Iteration Method (AIM)
\cite{cs,BHS,ff,hs,ns,ba,bam,br,bb,sh}.  In cases where such
polynomial solutions are not possible, AIM can also be used to
construct approximate wave functions and eigenvalues.  We
summarize AIM in the next section, and we formulate the eigenvalue
problem for general $q \ge 1$ in section~3.  In sections 4 and 5
we discuss in detail the cases $q=1$ and $q = 2$, respectively,
providing tables of exact analytical solutions, where possible,
and approximations in other cases.
\section{The asymptotic iteration method}
\noindent
The asymptotic iteration method (AIM) was first introduced as a technique capable of overcoming the
computation complexity of solving Schr\"odinger's equation  with singular potentials. It was first used as an
approximation method for computing both eigenvalues and eigenstates, with the aid of a computer-algebra system.
Because of its explicit iterative structure, the the method  has subsequently been employed very effectively to
obtain exact analytic solutions to many problems involving linear differential equations, especially those
presented by non-relativistic and relativistic quantum mechanics. Recently, some progress has been made also in
adapting the method to non-linear differential equations \cite{ss}. Perhaps because of its success in finding
exact solutions, the strength of AIM as an approximation method  has not been emphasized. For useful approximate
solutions, the iteration sequence must be started with a suitable form for the wave function, and also a starting
position $r = r_0$.  If these two seeds are well chosen, then either the sequence stops at an exact solution, or a
non-ending iteration sequence is generated, which can then be terminated to yield a good approximation.
Asymptotics usually provides a reliable guide to the initial form for the wave function; the starting point
$r_0$ is then chosen so as to stabilize the resulting iteration process. The present work illustrates the
effectiveness of AIM in finding both exact and approximate solutions of quantum-mechanical problems
generated by soft-core Coulomb potentials.  We shall now outline the method. Consider a differential
equation of the form
\begin{equation}\label{sec2eq1}
y''=\lambda_0(r) y'+s_0(r) y,\quad\quad ({}^\prime={d\over dr})
\end{equation}
where $\lambda_0(r)$ and $s_0(r)$ are $C^{\infty}-$differentiable
functions. A key feature of this method is to note the invariant
structure of the right-hand side of (\ref{sec2eq1}) under further
differentiation. Indeed, if we differentiate (\ref{sec2eq1}) with
respect to $r$, we obtain
\begin{equation}\label{sec2eq2}
y^{\prime\prime\prime}=\lambda_1 y^\prime+s_1 y
\end{equation}
where $\lambda_1= \lambda_0^\prime+s_0+\lambda_0^2$ and
$s_1=s_0^\prime+s_0\lambda_0.$ If we find the second derivative of
equation (\ref{sec2eq1}), we obtain
\begin{equation}\label{sec2eq3}
y^{(4)}=\lambda_2 y^\prime+s_2 y
\end{equation}
where $\lambda_2= \lambda_1^\prime+s_1+\lambda_0\lambda_1$ and
$s_2=s_1^\prime+s_0\lambda_1.$ Thus, for $(n+1)^{th}$ and
$(n+2)^{th}$ derivative of (\ref{sec2eq1}), $n=1,2,\dots$, we have
\begin{equation}\label{sec2eq4}
y^{(n+1)}=\lambda_{n-1}y^\prime+s_{n-1}y
\end{equation}
and
\begin{equation}\label{sec2eq5}
y^{(n+2)}=\lambda_{n}y^\prime+s_{n}y
\end{equation}
respectively, where
\begin{equation}\label{sec2eq6}
\lambda_{n}=
\lambda_{n-1}^\prime+s_{n-1}+\lambda_0\lambda_{n-1}\hbox{ ~~and~~
} s_{n}=s_{n-1}^\prime+s_0\lambda_{n-1}.
\end{equation}
From (\ref{sec2eq4}) and (\ref{sec2eq5}) we have
\begin{equation}\label{sec2eq7}
\lambda_n y^{(n+1)}- \lambda_{n-1}y^{(n+2)} = \delta_ny {\rm
~~~where~~~}\delta_n=\lambda_n s_{n-1}-\lambda_{n-1}s_n.
\end{equation}

Clearly, from (\ref{sec2eq7}) if $y$, the solution of
(\ref{sec2eq1}), is a polynomial of degree $n$, then $\delta_n\equiv
0$. Further, if $\delta_n=0$, then $\delta_{n'}=0$ for all $n'\geq
n$. In an earlier paper \cite{cs} we proved the principal theorem
of the Asymptotic Iteration Method (AIM), namely
\vskip0.1in
\noindent{\bf Theorem 1:} \emph{Given $\lambda_0$ and $s_0$ in
$C^{\infty}(a,b),$ the differential equation (\ref{sec2eq1}) has the
general solution
\begin{equation}\label{sec2eq8}
y(r)= \exp\left(-\int\limits^{r}\alpha(t) dt\right) \left[C_2
+C_1\int\limits^{r}\exp\left(\int\limits^{t}(\lambda_0(\tau) +
2\alpha(\tau)) d\tau \right)dt\right]
\end{equation}
if for some $n>0$
\begin{equation}\label{sec2eq9}
{s_{n}\over \lambda_{n}}={s_{n-1}\over \lambda_{n-1}} \equiv
\alpha.
\end{equation}
}

\noindent
For a given potential, the radial Schr\"odinger equation is
converted to the form of equation (\ref{sec2eq1}).
Then, $s_0(r)$ and $\lambda_0(r)$
are determined and the $s_n(r)$ and $\lambda_n(r)$ parameters are calculated
from the recurrence relations given by equation (\ref{sec2eq6}). The energy
eigenvalues are obtained from the roots of the termination condition of the method
in equation (\ref{sec2eq9}), or equivalently,
\begin{equation}\label{sec2eq10}
\delta_n(E;r)=\lambda_n(E;r) s_{n-1}(E;r)-\lambda_{n-1}(E;r)s_n(E;r)\equiv 0, \quad n=1,2,\dots.
\end{equation}
where $n$ represents the iteration number. For the exactly solvable
problems, the energy eigenvalues are obtained immediately from the equation
$$\delta_n(E;r)=\delta_n(E)\equiv 0,$$
and very often the eigenvalue index is the same as the iteration
number $n$. For problems that do  not  have exact solutions
expressible in the form chosen,
 the approximations for the eigenvalues are taken to be the roots of this same equation for large
iteration number.
\section{Soft-core Coulomb Potentials $V_q(r)$}
\noindent In atomic units, the radial Schr\"odinger equation for the potential $V_q(r)$ reads
\begin{equation}\label{sec3eq1}
 \left[-\half{d^2\over dr^2} + {l(l+1)\over 2r^2}-\frac{Z}{(r^q+\beta^q)^{\frac{1}{q}}}\right]\psi=E\psi.
\end{equation}
We may assume the solution of equation (\ref{sec3eq1}), which vanishes at the origin and at infinity, is
\begin{equation}\label{sec3eq2}
 \psi(r)=r^{l+1} e^{-k(r^q+\beta^q)^{\frac{1}{q}}}f((r^q+\beta^q)^{\frac{1}{q}})
\end{equation}
Straightforward computation shows that $f((r^q+\beta^q)^{\frac{1}{q}})$ is a solution of the
second-order homogeneous linear differential equation
\begin{align}\label{sec3eq3}
r^{2q-2}(r^q+\beta^q)^{\frac{2}{q}-2}f''((r^q+\beta^q)^{\frac{1}{q}})&+
\bigg[(q-1)\beta^qr^{q-2}(r^q+\beta^q)^{\frac{1}{q}-2}\notag\\
&+2\nu r^{q-2}(r^q+\beta^q)^{\frac{1}{q}-1}-2kr^{2q-2}(r^q+\beta^q)^{\frac{2}{q}-2}\bigg]f'((r^q+\beta^q)^{\frac{1}{q}})\notag\\
&+\bigg[-k(q-1)\beta^q r^{q-2}(r^q+\beta^q)^{\frac{1}{q}-2}-2k\nu r^{q-2}(r^q+\beta^q)^{\frac{1}{q}-1}\notag\\
&+k^2r^{2q-2}(r^q+\beta^q)^{\frac{2}{q}-2}+2Z(r^q+\beta^q)^{-\frac{1}{q}}+2E\bigg]f((r^q+\beta^q)^{\frac{1}{q}})=0.
\end{align}
where, for simplicity, we write $\nu=l+1$. From equation (\ref{sec3eq3}), we obtain, after some simplification,
\begin{align}\label{sec3eq4}
f''((r^q+\beta^q)^{\frac{1}{q}})&+
\bigg[{(q-1)\beta^q\over r^{q}(r^q+\beta^q)^{\frac{1}{q}}}+{2\nu\over r^{q}(r^q+\beta^q)^{\frac{1}{q}}}-2k\bigg]f'((r^q+\beta^q)^{\frac{1}{q}})\notag\\
&+\bigg[-{k(q-1)\beta^q\over r^{q}(r^q+\beta^q)^{\frac{1}{q}}}-{2k\nu\over r^{q}(r^q+\beta^q)^{\frac{1}{q}-1}}+k^2+{2Z\over r^{2q-2}(r^q+\beta^q)^{\frac{3}{q}-2}}+{2E\over r^{2q-2}(r^q+\beta^q)^{\frac{2}{q}-2}}\bigg]f((r^q+\beta^q)^{\frac{1}{q}})=0.
\end{align}
If we denote
$
\chi=(r^q+\beta^q)^{\frac{1}{q}}
$, we may now write (\ref{sec3eq4}) in a more compact form, as
\begin{align}\label{sec3eq5}
f''(\chi)&+
\bigg[{(q-1)\beta^q\over \chi(\chi^q-\beta^q)}+{2\nu\over \chi^{1-q}(\chi^q-\beta^q)}-2k\bigg]f'(\chi)\notag\\
&+\bigg[-{k(q-1)\beta^q\over \chi(\chi^q-\beta^q)}-{2k\nu\over \chi^{1-q}(\chi^q-\beta^q)}+k^2+{2Z\over \chi^{3-2q}(\chi^q-\beta^q)^{2-\frac{2}{q}}}+{2E\over \chi^{2-2q}(\chi^q-\beta^q)^{2-\frac{2}{q}}}\bigg]f(\chi)=0,
\end{align}
where differentiation is now with respect to $\chi$. This equation is now in a form suitable for the application of AIM with:
\begin{equation}\label{sec3eq6}
\left\{ \begin{array}{l}
 \lambda_0(r)\equiv -{(q-1)\beta^q\over \chi(\chi^q-\beta^q)}-{2\nu\over \chi^{1-q}(\chi^q-\beta^q)}+2k,\\ \\
 s_0(r)\equiv {k(q-1)\beta^q\over \chi(\chi^q-\beta^q)}+{2k\nu\over \chi^{1-q}(\chi^q-\beta^q)}-k^2-{2Z\over \chi^{3-2q}(\chi^q-\beta^q)^{2-\frac{2}{q}}}-{2E\over \chi^{2-2q}(\chi^q-\beta^q)^{2-\frac{2}{q}}}.
       \end{array} \right.
\end{equation}
In the next two sections we shall study the most important cases for physical applications, namely $q=1$ and $q=2$.
\section{Bound states of shifted Coulomb potential $V_1(r)=-Z/(r+\beta)$}
\noindent The problem of determining the energy eigenvalues and eigenstates of the shifted Coulomb potential
\begin{equation}\label{sec4eq1}
V_1(r)=-{Z\over r+\beta},\quad \beta>0
\end{equation}
has been of some interest in the past. As mentioned in the introduction, the potential may serve as an
approximation to the potential due to a smeared charge distribution, rather than a point charge,
and may be appropriate for describing mesonic atoms \cite{ray}. Using equation (\ref{sec3eq2}),
the solution of Schr\"odinger equation with  the shifted Coulomb potential $V_1(r)$ is
\begin{equation}\label{sec4eq2}
 \psi(r)=r^{l+1} e^{-k\chi}f(\chi),\quad\quad \chi=r+\beta,
\end{equation}
where $f(\chi)$ is the solution of differential equation
\begin{align}\label{sec4eq3}
f''(\chi)&+
\bigg[{2\nu\over \chi-\beta}-2k\bigg]f'(\chi)+\bigg[-{2k\nu\over \chi-\beta}+k^2+{2Z\over \chi}+2E\bigg]f(\chi)=0
\end{align}
or, for $k^2=-2E$, we have
\begin{align}\label{sec4eq4}
f''(\chi)&+
\bigg[{2\nu\over \chi-\beta}-2k\bigg]f'(\chi)+\bigg[-{2k\nu\over \chi-\beta}+{2Z\over \chi}\bigg]f(\chi)=0.
\end{align}
By comparing this equation with equation (\ref{sec3eq6}), we can easily
write the initial $\lambda_0(r)$ and $s_0(r)$ values,
\begin{equation}\label{sec4eq5}
\left\{ \begin{array}{l}
 \lambda_0(r)\equiv -{2\nu\over \chi-\beta}+2k,\\ \\
 s_0(r)\equiv {2k\nu\over \chi-\beta}-{2Z\over \chi}
       \end{array} \right.
\end{equation}
By means of AIM sequences, Eq. (\ref{sec2eq6}), we may calculate $\lambda_n(r)$ and
$s_n(r)$. Equation (\ref{sec2eq10}) then gives
\begin{equation}\label{sec4eq6}
\delta_1\equiv 0\Rightarrow   (Z-k\nu)(Z-k\nu-k)\chi^2-2Z\beta(Z-k\nu-k)\chi+Zb(Z\beta-k\beta-\nu)\equiv 0.
\end{equation}
From this equation, for $k={Z/(\nu+1)}$, we have a condition on the parameter $\beta$, namely $Z\beta =\nu+1$.
In general, we have in terms of the iteration number $n$ (or the degree of the polynomial solution $f(\chi)$ of (\ref{sec4eq4}))
\begin{equation}\label{sec4eq7}
\delta_n\equiv 0\Rightarrow k={Z\over \nu+n},\quad\quad n=1,2,\dots,
\end{equation}
with conditions on the parameter $\beta$ reported earlier \cite{HSSC09}. For other values
of $k$ and $\beta$, not necessary obeying these conditions, the eigenvalues are then
computed by means of the termination condition (\ref{sec2eq10}), namely $\delta_n (r; E) = 0$.
As we mentioned above, the computation of the eigenvalues by means of (\ref{sec2eq10}) should be
independent of the choice of $r$.  However, for certain values of $r_0$, we may encounter
oscillations of the computed roots and values that seem to diverge, presumably owing to
rounding and computational errors in the algorithms used.  In practice, this problem is
avoided by re-choosing $r_0$. In Table~\ref{tab:table1}, we show some results of these
calculation obtained by means of the computer algebra system {\it Maple.} This environment
allows numerical calculation of the roots of equation (\ref{sec2eq10}) with arbitrary chosen precision.
In order to accelerate the computation we have written our own root-finding algorithm instead of
using the default procedure {\tt Solve} of {\it Maple}.
In Table~\ref{tab:table1}, we also report the iteration number along with the initial $r_0$ used to
obtain the eigenvalues accurate to the number of decimal places recorded. For much smaller
values of the parameter $\beta$, a large number of iterations are usually needed, along with careful adjustment of $r_0$.
\medskip

\begin{table}[http]
\caption{\label{tab:table1} Approximate solutions of the radial
Schr\"odinger equation (\ref{sec3eq1}) for $q=1$.}
\begin{ruledtabular}
\begin{tabular}{ccccc}
 $\beta$\quad& $l$& Energy& $N$& $r_0$\\
\hline
$200$&$0$& -0.003~653~168~9& 14& 210\\
&$1$& -0.003~169~532~8& 17& 210\\
&$2$& -0.002~798~561~8& 17& 210\\
&$3$& -0.002~498~271~8& 17& 210\\
\hline
$100$&$0$& -0.006~742~076~7& 17& 120\\
&$1$& -0.005~633~737~9& 19& 120\\
&$2$& -0.004~812~493~4& 23& 140\\
&$3$& -0.004~168~784~4& 13& 150\\
\hline
$50$&$0$& -0.012~194~692~6& 16& 75\\
&$1$& -0.009~717~588~4& 18& 100\\
&$2$& -0.007~962~796~3& 13& 100\\
&$3$& -0.006~643~881~8& 16& 110\\
\hline
$35$&$0$& -0.016~388~672~4& 20& 65\\
&$1$& -0.012~685~924~3& 16& 65\\
&$2$& -0.010~135~716~7& 12& 75\\
&$3$& -0.008~268~680~7& 14& 90\\
\hline
$20$&$0$& -0.025~669~937~8& 38& 65\\
&$1$& -0.018~846~220~6& 30& 65\\
&$2$& -0.014~387~207~9& 22& 65\\
&$3$& -0.011~278~293~3& 22& 75\\
\hline
$10$&$0$& -0.043~438~719~3& 56& 45\\
&$1$& -0.029~446~515~7& 49& 55\\
&$2$& -0.021~024~301~6& 43& 65\\
&$3$& -0.015~576~600~1& 42& 75\\
\end{tabular}
\end{ruledtabular}
\end{table}

\subsection{Analytic solutions for $V_1(r) = -Z/(r+ \beta)$ and the confluent Heun equation}
\noindent It is interesting that the polynomial solutions of $f(\chi)$ of equation (\ref{sec4eq4})
are completely captured by AIM through the relations governed by $k$ and $\beta$ reported in Table~1.
In other words, no other polynomial solutions are possible. In order to confirm this claim, we now
express the solution of the differential equation (\ref{sec4eq4}) in terms of the confluent Heun function.
From Appendix~I, it is clear that the differential equation (\ref{sec4eq4}) is a special case  of the
confluent Heun differential equation (\ref{a1}). Indeed with simple transformation $z={\chi/ \beta}$,
equation (\ref{sec4eq4}) can be written as
\begin{align}\label{sec4eq8}
{d^2 f\over dz^2}&+
\bigg[{2\nu\over z-1}-2k\beta \bigg]{df\over dz}+\bigg[-{2k\nu\beta\over z-1}+{2Z\beta\over z}\bigg]f=0.
\end{align}
Further, the transformation $z=1-t$ yields
\begin{align}\label{sec4eq9}
{d^2 f\over dt^2}&+
\bigg[{2\nu\over t}+2k\beta \bigg]{df\over dt}+\bigg[{2k\nu\beta\over t}-{2Z\beta\over t-1}\bigg]f=0,
\end{align}
which is clearly a special case of (\ref{a1}).  Exact solutions are given (see equations (\ref{a2}) and (\ref{a3}))
by
\begin{equation}\label{sec4eq9}
f(t)=He(2k\beta,2v-1,-1,-2Z\beta,{1\over 2},t).
\end{equation}
For the polynomial solutions of $f(t)$, the parameters of
$He(2k\beta,2v-1,-1,-2Z\beta,{1\over 2},t)$ have to obey the conditions
of equation (\ref{a4}). The first condition of (\ref{a4}) immediately yields
\begin{equation}\label{sec4eq9}
k={Z\over \nu+N}
\end{equation}
where $N$ is the degree of the polynomial solution. This is in complete
agreement with the results of AIM as given by (\ref{sec4eq7}). While the
second condition of (\ref{a4}) now yields the conditions under which these
polynomial solutions are possible, namely,
\begin{equation}\label{sec4eq9}
\Delta_{N+1}(2k\nu\beta)=0,
\end{equation}
where $\Delta_{N+1}(2k\nu\beta)$ is given by means of the tri-diagonal determinant
of Table \ref{tab:table2}. Since $t=1-z=1-{\chi\over \beta}=1-{r+\beta\over \beta}=-{r\over \beta}$, the Taylor series expansion $f_N(t)$ becomes
\begin{equation}\label{sec4eq10}
f_N(t)\equiv He\left({2Z\beta\over N+\nu},2\nu -1,-1,-2Z\beta,{1\over 2},-
{r\over \beta}\right)=\sum_{n=0}^\infty \nu_n\left({2Z\beta\over N+\nu},2\nu-1,-1,-2Z\beta,
{1\over 2}\right)\left(-{r\over \beta}\right)^n.
\end{equation}
Therefore, we have
\begin{align}\label{sec4eq11}
f_1(r)&=1+{Z r\over 1+\nu}
\end{align}
for
\begin{align}\label{sec4eq12}
\Delta_2\left({2Z\nu\beta\over \nu+1}\right)\equiv  Z\beta-\nu-1=0
\end{align}
and
\begin{align}\label{sec4eq13}
f_2(r)&=1+{Zr\over 2+\nu}+{Z(-\nu^2+(Z\beta-4)\nu+Z\beta-4)r^2\over \beta(\nu+2)^2(2\nu+1)}
\end{align}
if
\begin{align}\label{sec4eq14}
\Delta_3\left({2Z\nu\beta\over \nu+2}\right)\equiv Z^2(\nu+1)\beta^2-3Z(\nu+2)(\nu+1)\beta+(2\nu+1)(\nu+2)^2=0.
\end{align}
Further
\begin{align}\label{sec4eq15}
f_3(r)&=1+{Z r\over 3+\nu}+{Z(-\nu^2+(Z\beta-6)\nu+Z\beta-9)r^2\over \beta(\nu+3)^2(2\nu+1)}\notag\\
&+{Z(2\nu^4+(19-3Z\beta)\nu^3+(Z^2\beta^2-21Z\beta+63)\nu^2+(3Z^2\beta^2-45Z\beta+81)\nu+2Z^2\beta^2-27Z\beta+27)r^3\over \beta^2(6\nu+3)(\nu+3)^3(\nu+1)}
\end{align}
if
\begin{align}\label{sec4eq16}
\Delta_4\left({2Z\nu\beta\over \nu+3}\right)&\equiv Z^3(\nu+2)(\nu+1)\beta^3-6Z^2(\nu+3)(\nu+2)(\nu+1)\beta^2+Z(11\nu^2+28\nu+15)(\nu+3)^2\beta\notag\\
&-3(\nu+1)(\nu+3)^3(2\nu+1)=0,
\end{align}
and
\begin{align}\label{sec4eq17}
f_4(r)&=1+{Z r\over 4+\nu}+{Z(-\nu^2+(Z\beta-8)\nu+Z\beta-16)r^2\over \beta(\nu+4)^2(2\nu+1)}\notag\\
&+{Z(2\nu^4+(25-3Z\beta)\nu^3+(Z^2\beta^2-27Z\beta+108)\nu^2+(3Z^2\beta^2-72Z\beta+176)\nu+2Z^2\beta^2-48Z\beta+64)t^3\over \beta^2(6\nu+3)(\nu+4)^3(\nu+1)}\notag\\
&+{Z r^4\over 6\beta^3(\nu+4)^4(2\nu+1)(\nu+1)(2\nu+3)}\left({(-6\nu^6+(11Z\beta-105)\nu^5+(-6Z^2\beta^2+163Z\beta-723)\nu^4+(918Z\beta+Z^3\beta^3}\right.\notag\\
&-2448-66Z^2\beta^2)\nu^3+(-252Z^2\beta^2-4128+2408Z\beta+6Z^3\beta^3)\nu^2+(-3072+2848Z\beta-384Z^2\beta^2+11Z^3\beta^3)\nu\notag\\
&-768+1152Z\beta+6Z^3\beta^3-192Z^2\beta^2))
\end{align}
if
\begin{align}\label{sec4eq18}
\Delta_5\left({2Z\nu\beta\over \nu+4}\right)&\equiv Z^4(\nu+3)(\nu+2)(\nu+1)\beta^4-10Z^3(\nu+4)(\nu+3)(\nu+2)(\nu+1)\beta^3+Z^2(35\nu^3+195\nu^2+328\nu+162)(\nu+4)^2\beta^2\notag\\
&-Z(50\nu^3+231\nu^2+313\nu+126)(\nu+4)^3\beta+6(2\nu+3)(2n+1)(\nu+1)(\nu+4)^4=0.
\end{align}

\noindent Similarly, for the higher polynomials, in general,
$$\Delta_{N+1}(2k\nu\beta)\equiv 0$$
where $\Delta_{N+1}(2k\nu\beta)$ is given by Table \ref{tab:table2}.

\begin{table}[http]\tiny
\centering
\caption{\label{tab:table2} The determinant $\Delta_{N+1}(2Z\beta )$. Here $k={Z/(N+\nu)}$}
\begin{tabular}{|ccccccccccccc|}
$2k\nu\beta$& &$2\nu$& &$0$& &$\dots$& &$0$& &$0$& & $0$\\
& & & & & & & & & & & & \\
$2Nk\beta$& &$2k\nu\beta-(2\nu-2k\beta)$& & $2(2\nu+1)$& &\dots& &$0$& &$0$& & $0$\\
& & & & & & & & & & & & \\
$0$& & $2(N-1)k\beta$ & & $2k\nu\beta-2(2\nu+1-2k\beta)$& & $\dots$& & $0$& & $0$& &$0$\\
& & & & & & & & & & & & \\
$\vdots$& & $\vdots$& & $\vdots$& & $\ddots$& & $\vdots$& &$\vdots$& & $\vdots$\\
& & & & & & & & & & & &\\
$0$& &$0$& &$0$ & &$\dots$& & $2k\nu\beta-(N-2)(N+2\nu-2k\beta-3)$& & $(N-1)(N+2\nu-2)$ & &$0$ \\
& & & & & & & & & & & &\\
$0$& &$0$& &$0$ & &$\dots$& & $4k\beta$& & $2k\nu\beta-(N-1)(N+2\nu-2k\beta-2)$ & &$N(N+2\nu-1)$ \\
& & & & & & & & & & & &\\
$0$& &$0$& &$0$ & &$\dots$& & $0$& & $2k\beta$ & &$2k\nu\beta-N(N+2\nu+2k\beta-1)$ \\
\end{tabular}
\end{table}

\noindent All the $N+1$ roots of this determinant are real and distinct \cite{arscott}. It may be possible in future to obtain such results by studying the correspondence between the iteration details of AIM and the structure of the resulting tri-diagonal determinant.

\section{Bound states for the soft-core Coulomb potential $V_2(r)=-Z/\sqrt{r^2+\beta^2}$}
\noindent For $q=2$ and $\nu=0$, we have using Eq.(\ref{sec3eq5})
\begin{align}\label{sec3eq11}
f''(\chi)&+
\bigg[{\beta^2\over \chi(\chi^2-\beta^2)}+{2\nu\chi\over \chi^2-\beta^2}-2k\bigg]f'(\chi)+\bigg[-{k\beta^2\over \chi(\chi^2-\beta^2)}-{2k\nu\chi\over \chi^2-\beta^2}+k^2+{2Z\chi\over \chi^2-\beta^2}+{2E\chi^2\over \chi^2-\beta^2}\bigg]f(\chi)=0.
\end{align}
From this equation we have
\begin{align}\label{sec3eq12}
(1-\beta^2\chi^{-2})f''(\chi)&+
\bigg[\beta^2\chi^{-3}-2k+2k\beta^2\chi^{-2}\bigg]f'(\chi)+\bigg[-k\beta^2\chi^{-3}+k^2+
2Z\chi^{-1}+2E-k^2\beta^2\chi^{-2}\bigg]f(\chi)=0,
\end{align}
as given earlier by Liu \& Clark \cite{CWC} for $-2E=k^2$. For the purpose of applying AIM, we may note for $\nu=l+1$ and $$2E=-k^2$$ that
\begin{align}\label{sec3eq13}
f''(\chi)&=
\bigg[2k-{\beta^2\over \chi(\chi^{2}-\beta^2)}-{2\nu\chi\over \chi^2-\beta^2}\bigg]f'(\chi)+\bigg[{k\beta^2\over\chi(\chi^2-\beta^2)}+{2k\nu\chi\over \chi^2-\beta^2}+{k^2\beta^2\over \chi^2-\beta^2}-{2Z\chi\over \chi^2-\beta^2} \bigg]f(\chi).
\end{align}
This equation can be further simplified, yielding
\begin{align}\label{sec3eq14}
f''(\chi)&=
\bigg[2k+{1\over \chi}-{(2v+1)\chi\over \chi^2-\beta^2}\bigg]f'(\chi)+\bigg[-{k\over \chi}+{((2\nu+1)k-2Z)\chi+k^2\beta^2\over \chi^2-\beta^2} \bigg]f(\chi)
\end{align}
or
\begin{align}\label{sec3eq15}
f''(\chi)&=
\bigg[2k+{1\over \chi}-{(v+{1/2})\over \chi-\beta}-{(v+{1/2})\over \chi+\beta}\bigg]f'(\chi)+\bigg[{(\nu+1/2)k-Z+k^2\beta/2\over \chi-\beta}+ {(\nu+1/2)k-Z-k^2\beta/2\over \chi+\beta}-{k\over \chi}\bigg]f(\chi).
\end{align}
Thus we may now apply AIM with
\begin{equation}\label{sec3eq16}
\left\{\begin{array}{l}
 \lambda_0=2k+{1\over \chi}-{(v+{1/2})\over \chi-\beta}-{(v+{1/2})\over \chi+\beta},\\ \\
 s_0={(\nu+1/2)k-Z+k^2\beta/2\over \chi-\beta}+ {(\nu+1/2)k-Z-k^2\beta/2\over \chi+\beta}-{k\over \chi}.
       \end{array} \right.
\end{equation}
Using AIM sequence (\ref{sec2eq6}) we can thus produce a set of exact polynomial solutions given by
\begin{equation}\label{sec3eq17}
k={Z\over \nu+n}\Rightarrow E_{nl}=-{1\over 2}{Z^2\over (n+\nu)^2},
\end{equation}
where $n=1,2,3,\dots$ is the iteration number used by AIM, along
with the corresponding conditions on the potential parameter $\beta$ reported in
Table \ref{tab:table3}.

\begin{table}[http]
\caption{\label{tab:table3} Polynomial conditions on the potential parameters for the existence of exact solutions of the radial
Schr\"odinger equation (\ref{sec3eq15}) for $q=2$.}
\begin{ruledtabular}\footnotesize
\begin{tabular}{l|l}
 $k$& Conditions on $\beta$\\
\hline
$\frac{Z}{\nu+1}$& $Z^2\beta^2-2(\nu+1)^3=0$ \\
\\
$\frac{Z}{\nu+2}$& $Z^4\beta^4-6Z^2(\nu+2)^3\beta^2+4(2\nu+1)(\nu+2)^5=0$ \\
\\
$\frac{Z}{\nu+3}$& $Z^6\beta^6-12Z^4(\nu+3)^3\beta^4+4Z^2(11\nu+18)(\nu+3)^5\beta^2-24(2\nu+1)(\nu+1)(\nu+3)^7=0$ \\
\\
$\frac{Z}{\nu+4}$& $Z^8\beta^8-20Z^6(\nu+4)^3\beta^6+20Z^4(7\nu+19)(\nu+4)^5\beta^4-8Z^2(50\nu^2+193\nu+182)(\nu+4)^7\beta^2+96(2\nu+3)(2\nu+1)(\nu+1)(\nu+4)^9=0$\\
\\
$\frac{Z}{\nu+5}$& $Z^{10}\beta^{10}-30Z^8(\nu+5)^3\beta^8+20Z^6(17\nu+64)(\nu+5)^5\beta^6-24Z^4(75\nu^2+456\nu+685)(\nu+5)^7\beta^4$\\
&$+16Z^2(274\nu^3+1821\nu^2+3942\nu+2765)(\nu+5)^9\beta^2-960(2\nu+3)(2\nu+1)(\nu+2)(\nu+1)(\nu+5)^{11}=0$\\
\\
$\frac{Z}{\nu+6}$& $Z^{12}\beta^{12}-42Z^{10}(\nu+6)^3\beta^{10}+140Z^8(5\nu+24)(\nu+6)^5\beta^8-168Z^6(35\nu^2+288\nu+588)(\nu+6)^7\beta^6$\\
&$+16Z^4(1624\nu^3+16338\nu^2+54099\nu+58860)(\nu+6)^9\beta^4-288Z^2(196\nu^4+1956\nu^3+7157\nu^2+11330\nu+6528)(\nu+6)^{11}b^2$\\
&$+5760(2\nu+5)(2\nu+3)(2\nu+1)(\nu+2)(\nu+1)(\nu+6)^{13}=0$\\
\\
$\frac{Z}{\nu+7}$&$Z^{14}\beta^{14}-56Z^{12}(\nu+7)^3\beta^{12}+56Z^{10}(23\nu+134)(\nu+7)^5\beta^{10}-112Z^8(140\nu^2+1447\nu+3719)(\nu+7)^7\beta^8$\\
&$+16Z^6(6769\nu^3+90363\nu^2+398730\nu+581140)(\nu+7)^9\beta^6-64Z^4(6566\nu^4+96071\nu^3+520284\nu^2+1234091\nu+108054)(\nu+7)^{11}\beta^4$\\
&$+576Z^2(1452\nu^5+20112\nu^4+109039\nu^3+288179\nu^2+370287\nu+184667)(\nu+7)^{13}\beta^2$\\
&$-80640(2\nu+5)(2\nu+3)(2\nu+1)(\nu+3)(\nu+2)(\nu+1)(\nu+7)^{15}=0$\\
\end{tabular}
\end{ruledtabular}
\end{table}

\vskip0.1 true in
\noindent The asymptotic iteration method can also be used to compute approximate eigenvalues of Schr\"odinger equation  (\ref{sec3eq15})
for different explicit values of the potential parameters $Z$ and $
\beta$.  We use $\lambda_0$ and $s_0$ given by (\ref{sec3eq16}), and the numerical results,
including the iteration number and suitable $r_0$, are reported in Table (\ref{tab:table4}).
It is important to note for the case of $\beta=200$, the eigenvalues for $l=0,1,2,3$ are
squeezed together in space less than $0.001$, and this makes it hard for AIM to function correctly.
In such cases, it is therefore recommended to use the scaling law as given by (\ref{sec1eq2}).
For example, we know using the scaling law to show that
$$E(1,200,2)={1\over 4^2}E(4,50,2).$$
Thus instead of computing $E(1,200,2)$ we may compute $E(4,50,2)$ then divide the result by $16$.
Indeed, for $l=2$ and $l=3$, AIM yields for $E(4,40,2)$, the values $-0.062~403~136~6$ and $-0.057~981~633~1,$ respectively.
From which we can now compute the eigenvalues $E(1,200,2)$ as reported in Table (\ref{tab:table4}).

\begin{table}
\caption{\label{tab:table4} Approximate solutions of the radial
Schr\"odinger equation (\ref{sec3eq1}) for $q=2$. The values marked $^\dagger$ are computed using a scaling law, as discussed in the text.}
\begin{ruledtabular}
\begin{tabular}{ccccc}
 $\beta$\quad& $l$& Energy& $N$& $r_0$\\
\hline
$200$&$0$& -0.004~502~854~6& 23& 3\\
&$1$& -0.004~193~071~3& 21& 3\\
&$2$& -0.003~900~196~0$^\dagger$& 16& 2\\
&$3$& -0.003~623~852~1$^\dagger$& 16& 2\\
\hline
$100$&$0$& -0.008~629~777~4& 21& 3\\
&$1$& -0.007~800~132~8& 18& 3\\
&$2$& -0.007~035~193~2& 21& 3\\
&$3$& -0.006~332~729~0& 19& 3\\
\hline
$50$&$0$& -0.016~260~721~2& 23& 3\\
&$1$& -0.014~088~375~1& 19& 3\\
&$2$& -0.012~158~711~5& 22& 3\\
&$3$& -0.010~458~421~2& 19& 3\\
\hline
$35$&$0$& -0.022~334~284~2& 23& 3\\
&$1$& -0.018~810~954~4& 23& 3\\
&$2$& -0.015~761~480~4& 22& 3\\
&$3$& -0.013~152~440~8& 21& 3\\
\hline
$20$&$0$& -0.036~198~545~7& 25& 3\\
&$1$& -0.028~830~115~2& 24& 3\\
&$2$& -0.022~787~158~5& 23& 3\\
&$3$& -0.017~928~611~8& 24& 3\\
\hline
$10$&$0$& -0.063~738~918~2& 27& 3\\
&$1$& -0.046~199~039~0& 27& 3\\
&$2$& -0.033~158~588~8& 32& 4\\
&$3$& -0.023~806~736~2& 37& 5\\
\end{tabular}
\end{ruledtabular}
\end{table}

\subsection{Analytic solutions for $V_2(r) = -{Z/\sqrt{r^2+ \beta^2}}$ and the generalized Heun equation}
\noindent To analyze the analytic solutions of the differential equation (\ref{sec3eq15}), we first let $\chi=\beta z$. We then have
\begin{align}\label{sec3eq18}
f''&=
\bigg[2k\beta+{1\over z}-{(v+{1/2})\over z-1}-{(v+{1/2})\over z+1}\bigg]f'+\bigg[{(\nu+1/2)k\beta-Z\beta+k^2\beta^2/2\over z-1}+ {(\nu+1/2)k\beta-Z\beta-k^2\beta^2/2\over z+1}-{k\beta\over z}\bigg]f
\end{align}
or,  in the more appropriate form,
\begin{align}\label{sec3eq19}
f''&=
\bigg[2k\beta+{1\over z}-{(v+{1/2})\over z-1}-{(v+{1/2})\over z+1}\bigg]f'+\bigg[{\beta k+\beta^2k^2z+2\beta(k\nu-Z)z^2\over z(z-1)(z+1)}
\bigg]f.
\end{align}
This equation can be regarded as a special case of the generalized Heun differential equation.
According to Sch\"afke and Schmidt \cite{schafke}, the generalized Heun equation has the form
\begin{align}\label{sec3eq20}
f''+
\bigg[\alpha+{1-\mu_0\over z}+{1-\mu_1\over z-1}+{1-\mu_2\over z-\mbox{\^{a}}}\bigg]f'+
\bigg[{\beta_0+\beta_1z+\beta_2z^2\over z(z-1)(z-\mbox{\^{a}})}
\bigg]f=0
\end{align}
in which \^{a} $\in C\backslash\{0,1\}$ and $\mu_0,\mu_1,\mu_2,\alpha,\beta_0,\beta_1,\beta_2$ are arbitrary complex numbers.
For $\alpha\neq 0$, the generalized Heun equation (\ref{sec3eq20}) has three regular singular points, at $0,1,$ \^{a}  with
exponents (i.e. the roots of the indicial equations) $\{0,\mu_0\}$, $\{0,\mu_1\}$ and $\{0,\mu_2\}$, respectively,
in addition to one irregular singular point at infinity. These can be shown to be given by:
\begin{enumerate}
  \item For the singularity $z=0$, we have
  \begin{align*}
    \lim\limits_{z\rightarrow 0} z\left(\alpha+{1-\mu_0\over z}+{1-\mu_1\over z-1}+{1-\mu_2\over z-\mbox{\^{a}}}\right)&=1-\mu_0\\
\lim\limits_{z\rightarrow 0} z^2\left({\beta_0+\beta_1 z+\beta_2 z^2\over z(z-1)(z-\mbox{\^{a}})}\right)&=0
\end{align*}
and the indicial equation then reads
$$s(s-1)+(1-\mu_0)s=0\Rightarrow s(s-\mu_0)=0$$
with simple roots $\{0,\mu_0\}$.
\item For the singularity $z=1$, we have
  \begin{align*}
    \lim\limits_{z\rightarrow 1} (z-1)\left(\alpha+{1-\mu_0\over z}+{1-\mu_1\over z-1}+{1-\mu_2\over z-\mbox{\^{a}}}\right)&=1-\mu_1\\
\lim\limits_{z\rightarrow 1} (z-1)^2\left({\beta_0+\beta_1 z+\beta_2 z^2\over z(z-1)(z-\mbox{\^{a}})}\right)&=0
\end{align*}
and the indicial equation then reads
$$s(s-1)+(1-\mu_1)s=0\Rightarrow s(s-\mu_1)=0$$
with simple roots $\{0,\mu_1\}$.
\item For the singularity $z=$\^{a}, we have
  \begin{align*}
    \lim\limits_{z\rightarrow \mbox{\^{a}}} (z-\mbox{\^{a}})\left(\alpha+{1-\mu_0\over z}+{1-\mu_1\over z-1}+{1-\mu_2\over z-\mbox{\^{a}}}\right)&=1-\mu_2\\
\lim\limits_{z\rightarrow \mbox{\^{a}}} (z-\mbox{\^{a}})^2\left({\beta_0+\beta_1 z+\beta_2 z^2\over z(z-1)(z-\mbox{\^{a}})}\right)&=0
\end{align*}
and the indicial equation then reads
$$s(s-1)+(1-\mu_2)s=0\Rightarrow s(s-\mu_2)=0$$
with simple roots $\{0,\mu_2\}$.
\end{enumerate}
To find the exponent of the singular point at $\infty$, we substitute $z={1\over z_1}$, $z_1={1\over z}$, $dz_1=-{1\over z^2}dz$. Further
$${df\over dz}={df\over dz_1}{dz_1\over dz}=-{1\over z^2}{df\over dz_1}=-z_1^2{df\over dz_1}$$
and
$${d^2f\over dz^2}={d\over dz}{df\over dz}={d\over dz_1}\left(-z_1^2{df\over dz_1}\right){dz_1\over dz}=\left(-2z_1 {df\over dz_1}-z_1^2{d^2f\over dz_1^2}\right)\left(-z_1^2\right)=2z_1^3{df\over dz_1}+z_1^4{d^2f\over dz_1^2}$$
Substituting this in (\ref{sec3eq20}), we obtain
\begin{align}
2z_1^3{df\over dz_1}+z_1^4{d^2f\over dz_1^2}+\left[\alpha+(1-\mu_0)z_1+z_1\left({1-\mu_1\over 1-z_1}\right)+z_1{1-\mu_0\over 1-\mbox{\^{a}}z_1}\right]\left(-z_1^2{df\over dz_1}\right)+
\left({\beta_0 z_1^2+\beta_1 z_1+\beta_2\over {z_1(1-z_1)(1-\mbox{\^{a}}z_1)}}\right)f=0.
\end{align}
Thus we have for $\alpha\neq 0$ that $z=\infty$ is an irregular singularity of Poincar\'e rank~$1$, since the term $z_1^{k+1}\left({\alpha/ z_1^2}\right)=\alpha\neq 0$ for $k=1$. Clearly then for the Heun equation (\ref{sec3eq20}) with $\alpha=0$, the singular points $0,1,\mbox{\^a}$ and  infinity are  regular singular points. This is the main difference between the Heun differential equation ($\alpha=0$) and the generalized Heun equation as given by (\ref{sec3eq20}).
\medskip

\noindent In order to find a power series solution of (\ref{sec3eq20}) about the singular point $z=0$, i.e.
valid for $|z|<1$, we notice first that equation (\ref{sec3eq20}) can be written as
\begin{align}\label{sec3eq21}
(z^3-(1+b)z^2+\mbox{\^{a}}z)f''&+
\bigg[\alpha z^3+(3-\alpha(1+\mbox{\^a})-\mu_0-\mu_1-\mu_2)z^2+(-2+\mu_0+\mu_2+(-2+\mu_0+\mu_1)\mbox{\^a}+\alpha\mbox{\^a})z\notag\\
&+(1-\mu_0)\mbox{\^a} \bigg]f'+
\bigg[\beta_0+\beta_1z+\beta_2z^2\bigg]f=0.
\end{align}
Using
$$f(z)=\sum_{k=0}^\infty c_kz^k,\quad {df\over dz}=\sum_{k=1}^\infty kc_kz^{k-1},\quad {d^2f\over dz^2}=\sum_{k=2}^\infty k(k-1)c_kz^{k-2}$$
and substitute in (\ref{sec3eq21}), we obtain, after some simplification,
\begin{align}\label{sec3eq22}
\sum_{k=3}^\infty &(k-1)(k-2)c_{k-1}z^k-(1+\mbox{\^a})\sum_{k=2}^\infty k(k-1)c_kz^k+\mbox{\^a}\sum_{k=1}^\infty k(k+1)c_{k+1}z^k+\alpha\sum_{k=3}^\infty (k-2)c_{k-2}z^k\notag\\
&+(3-\alpha(1+\mbox{\^a})-\mu_0-\mu_1-\mu_2)\sum_{k=2}^\infty (k-1)c_{k-1}z^k+(-2+\mu_0+\mu_2+(-2+\mu_0+\mu_1)\mbox{\^a}+\alpha\mbox{\^a})\sum_{k=1}^\infty kc_kz^k\notag\\
&+(1-\mu_0)\mbox{\^a}\sum_{k=0}^\infty (k+1)c_{k+1}z^k+\beta_0\sum_{k=0}^\infty c_{k}z^{k}+\beta_1\sum_{k=1}^\infty c_{k-1}z^{k-1}+B_2\sum_{k=2}^\infty c_{k-2}z^k=0
\end{align}
from which we obtain the four-term recurrence relation
\begin{align}\label{sec3eq23}
\left[(k-1)(k-2)+(3-\alpha(1+\mbox{\^a})-\mu_0-\mu_1-\mu_2)(k-1)+B_1\right]c_{k-1}\notag\\
+\left[-k(k-1)(1+\mbox{\^a})+k(-2+\mu_0+\mu_2+(-2+\mu_0+\mu_1)\mbox{\^a}+\alpha \mbox{\^a})+B_0\right]c_k\notag\\
+\left[\mbox{\^a}k(k+1)+\mbox{\^a}(k+1)(1-\mu_0)\right]c_{k+1}+\left[\alpha(k-2)+B_2\right]c_{k-2}=0.
\end{align}
Replacing $k$ by $k-1$, we obtain
\begin{align}\label{sec3eq24}
\mbox{\^a}k(k-\mu_0)c_k+\left[(k-1)\bigg(-(k-2)(1+\mbox{\^a})-2+\mu_0+\mu_2+(-2+\mu_0+\mu_1)\mbox{\^a}+\alpha\mbox{\^a}\bigg) +B_0\right]c_{k-1}\notag\\
+\bigg[(k-2)\bigg(k-\alpha(1+\mbox{\^a})-\mu_0-\mu_1-\mu_2\bigg)+B_1\bigg]c_{k-2}+\bigg[\alpha(k-3)+B_2\bigg]c_{k-3}=0
\end{align}
with $c_{-1}=c_{-2}=0$. It should be noted that the Floquet solutions of Sch\"afke and Schmidt \cite{schafke} to
the generalized Heun equation (\ref{sec3eq20}) follows by using
$$c_k={\tau_k(\mu_0,\mu_1,\mu_2,\alpha,B_0,B_1,B_2;\mbox{\^a})\over \Gamma(k+1-\mu_0)\Gamma(k+1)},\quad \tau_{-1}=\tau_{-2}=0$$
to obtain, using (\ref{sec3eq23}), the four-term recurrence relation
\begin{align}\label{sec3eq25}
\mbox{\^a}\tau_k&=\left[(k-1)\bigg(k-\mu_0-\mu_2+(k-\mu_0-\mu_1)\mbox{\^a}-\alpha\mbox{\^a}\bigg)-B_0\right]\tau_{k-1}\notag\\
&+\left[(k-2)\bigg(-k+\alpha(\mbox{\^a}+1)+\mu_0+\mu_1+\mu_2\bigg)-B_1\right](k-1)(k-1-\mu_0)\tau_{k-2}\notag\\
&-\left[\alpha(k-3)+B_2\right](k-2-\mu_0)(k-1-\mu_0)(k-1)(k-2)\tau_{k-3}.
\end{align}
That is to say, in the notation of Sch\"afke and Schmidt,
\begin{align}\label{sec3eq26}
\tau_k&=\phi_1(k-1)\tau_{k-1}-\phi_2(k-2)\tau_{k-2}+\phi_3(k-3)\tau_{k-3},
\end{align}
where
\begin{align}\label{sec3eq27}
\phi_1(\xi)&=\xi(\xi+1-\mu_0-\mu_1)+{1\over \mbox{\^a}}\xi(\xi+1-\mu_0-\mu_2)-\alpha\xi-{B_0\over \mbox{\^a}}\notag\\
\phi_2(\xi)&=(\xi+1)(\xi+1-\mu_0)\left({1\over\mbox{\^a}}\xi(\xi+2-\mu_0-\mu_1-\mu_2)-(1+{1\over \mbox{\^a}})\alpha \xi +{B_1\over \mbox{\^a}}\right)\notag\\
\phi_3(\xi)&=-{1\over \mbox{\^a}}(\alpha\xi+B_2)(\xi+1-\mu_0)(\xi+2-\mu_0)(\xi+1)(\xi+2).
\end{align}
The condition for an $N$-degree polynomial solution is equivalent to the equation  $c_{N+3}=0$ in (\ref{sec3eq24}), i.e.
\begin{equation}\label{sec3eq28}
\alpha N+B_2=0.
\end{equation}
If this condition is satisfied, the recurrence relation
\begin{align}\label{sec3eq29}
\mbox{\^a}(k+1)(k+1-\mu_0)c_{k+1}+\left[k\bigg(-(k-1)(1+\mbox{\^a})-2+\mu_0+\mu_2+(-2+\mu_0+\mu_1)\mbox{\^a}+\alpha\mbox{\^a}\bigg) +B_0\right]c_{k}\notag\\
+\bigg[(k-1)\bigg(k+1-\alpha(1+\mbox{\^a})-\mu_0-\mu_1-\mu_2\bigg)+B_1\bigg]c_{k-1}+\bigg[\alpha(k-2)+B_2\bigg]c_{k-2}=0
\end{align}
for $0\leq k\leq N$ forms a system of $N+1$ homogeneous linear relation in the $N+1$ coefficients $c_k$.
This system has a non-trivial solution if and only if its determinant $\Delta_{N+1}$, given explicitly in Table~V, is zero; that is to say,
\begin{equation}\label{sec3eq30}
\Delta_{N+1}=0.
\end{equation}
Under the simultaneous fulfillment of the two conditions (\ref{sec3eq28}) and (\ref{sec3eq30}),
the solution of the generalized Heun equation
(\ref{sec3eq20}) reduces to a polynomial of degree $N$. In what follows, we use these
conditions to compute the polynomial solutions of the differential equation (\ref{sec3eq19}).
These results confirm our earlier findings using AIM and reported in Table (\ref{tab:table3}).

\begin{sidewaystable}\tiny
\centering
\caption{\label{tab:table5} The determinant $\Delta_{N+1}$.}
\begin{tabular}{|ccccccccccccccc|}
$B_1$& &$-2+\mu_0+\mu_2+(-2+\mu_0+\mu_1+\alpha)\mbox{\^a}+B_0$& &$2\mbox{\^a}(2-\mu_0)$& &$0$& &$0$& &$0$& & $0$& &\dots\\
& & & & & & & & & & & & & &\\
$B_2$& &$(3-\alpha(1+\mbox{\^a})-\mu_0-\mu_1-\mu_2)+B_1$& & $2(-3+\mu_0+\mu_2+(-3+\mu_0+\mu_1+\alpha)\mbox{\^a})+B_0$& &$3\mbox{\^a}(3-\mu_0)$& &$0$& &$0$& & $0$& &\dots\\
& & & & & & & & & & & & & &\\
$0$& & $B_2+\alpha$ & & $2(4-\alpha(1+\mbox{\^a})-\mu_0-\mu_1-\mu_2)+B_1$& & $3(-4+\mu_0+\mu_2+(-4+\mu_0+\mu_1+\alpha)\mbox{\^a})+B_0$& & $4\mbox{\^a}(4-\mu_0)$& & $0$& &$0$& &\dots\\
& & & & & & & & & & & & & &\\
$0$& & $0$& & $B_2+2\alpha$& & $3(5-\alpha(1+\mbox{\^a})-\mu_0-\mu_1-\mu_2)+B_1$& & $4(-5+\mu_0+\mu_2+(-5+\mu_0+\mu_1+\alpha)\mbox{\^a})+B_0$& &$0$& & $0$& &\dots\\
& & & & & & & & & & & & & &\\
$\dots$& & $\dots$& & $\dots$& & $\dots$& & $\dots$& &$\dots$& & $\dots$& &\dots\\
\end{tabular}
\end{sidewaystable}
\medskip

\noindent Comparing Eq.(\ref{sec3eq19}) and Eq.(\ref{sec3eq20}), we have
\begin{align}\label{sec3eq31}
\mu_0&=2,\quad \mu_1={1\over 2}-\nu,\quad \mu_2={1\over 2}-\nu\notag\\
B_0&=-\beta k,\quad B_1=-\beta^2 k^2,\quad B_2=-2\beta(k\nu-Z)\notag\\
\alpha&=-2k\beta,\quad \mbox{\^a}=-1.
\end{align}

\noindent Condition (\ref{sec3eq28}), immediately yields
\begin{equation}\label{sec3eq32}
k={Z\over N+\nu}.
\end{equation}
The polynomial solutions can be found by means of the condition (\ref{sec3eq30}): for $N=1$, we have
\begin{equation}\label{sec3eq33}
\Delta_2=\left| \begin{array}{cc}
-\beta^2 k^2 & \beta k\\
-2\beta(k\nu-Z) & 2\nu-\beta^2 k^2 \\
\end{array}\right| = \beta^2 k(\beta^2k^3-2Z)=0.
\end{equation}
From this, for $k={Z/(\nu+1)}$, we have
$$Z^2\beta^2-2(\nu+1)^3=0$$
with polynomial solution given by
\begin{align}\label{sec3eq34}
f_1(z)&=1+{\beta Z\over \nu+1} z\notag\\
&=1+{Z\over \nu+1}\chi\notag\\
&=1+{Z\over \nu+1}\sqrt{r^2+\beta^2}.
\end{align}
For $N=2$, we have
\begin{equation}\label{sec3eq35}
\Delta_3=\left| \begin{array}{ccccc}
-\beta^2 k^2 & &\beta k& &0\\
-2\beta(k\nu-Z) & &2\nu-\beta^2 k^2 & &3k\beta\\
0& & -2\beta(k\nu-Z)-2\beta k& &2+4\nu-\beta^2 k^2
\end{array}\right| = 0
\end{equation}
from which,  for $k={Z/(\nu+2)}$, we have
$$Z^4\beta^4-6Z^2(\nu+2)^3\beta^2+4(2\nu+1)(\nu+2)^5=0.$$
with polynomial solution given by
\begin{align}\label{sec3eq36}
f_2(z)&=1+ {Z\beta\over \nu+2} z+{2\beta^2Z^2\over \beta^2 Z^2-2(2\nu+1)(2+\nu)^2}z^2\notag\\
&=1+{Z\over \nu+2}\chi+{2Z^2\over \beta^2 Z^2-2(2\nu+1)(2+\nu)^2}\chi^2\notag\\
&=1+{Z\over \nu+2}\sqrt{r^2+\beta^2}+{2Z^2\over \beta^2 Z^2-2(2\nu+1)(2+\nu)^2}(\sqrt{r^2+\beta^2})^2
\end{align}
For $N=3$, we have
\begin{equation}\label{sec3eq37}
\Delta_4=\left| \begin{array}{ccccccc}
-\beta^2 k^2 &  &\beta k& &0& & 0\\
-2\beta(k\nu-Z) & &2\nu-\beta^2 k^2 & &3k\beta& & -3\\
0& & -2\beta(k\nu-Z)-2\beta k& &2+4\nu-\beta^2 k^2& & 5\beta k\\
0& & 0& & -2\beta(k\nu-Z)-4\beta k& & 6+6\nu-\beta^2 k^2
\end{array}\right| = 0
\end{equation}
from which,  for $k={Z/(\nu+3)}$, we have
$$Z^6\beta^6-12Z^4(\nu+3)^3\beta^4+4Z^2(11\nu+18)(\nu+3)^5\beta^2-24(2\nu+1)(\nu+1)(\nu+3)^7=0.$$
with polynomial solution given by (note $k=Z/(\nu+3)$)
\begin{align}\label{sec3eq38}
f_3(z)&=1+ {\beta k} z-{2\beta^2k(\beta^2 k^2-6(\nu+1))(k(\nu+1)-Z)\over \beta^4k^4+2k(6k-5Z)\beta^2+12(\nu+1)(2\nu+1)}z^2+{4k\beta^3(k(1+\nu)-Z)(k(\nu+2)-Z)\over \beta^4 k^4+2k(6k-5Z)\beta^2+12(\nu+1)(2n+1)}z^3\notag\\
&=1+ k\chi-{2k(\beta^2 k^2-6(\nu+1))(k(\nu+1)-Z)\over \beta^4k^4+2k(6k-5Z)\beta^2+12(\nu+1)(2\nu+1)}\chi^2+{4k(k(1+\nu)-Z)(k(\nu+2)-Z)\over \beta^4 k^4+2k(6k-5Z)\beta^2+12(\nu+1)(2n+1)}\chi^3\notag\\
&=1+ k\sqrt{r^2+\beta^2}-{2k(\beta^2 k^2-6(\nu+1))(k(\nu+1)-Z)\over \beta^4k^4+2k(6k-5Z)\beta^2+12(\nu+1)(2\nu+1)}(\sqrt{r^2+\beta^2})^2\notag\\
&+{4k(k(1+\nu)-Z)(k(\nu+2)-Z)\over \beta^4 k^4+2k(6k-5Z)\beta^2+12(\nu+1)(2n+1)}(\sqrt{r^2+\beta^2})^3.
\end{align}
Similarly, higher degree polynomial solutions follow from the determinant $\Delta_{N+1}=0$ given by Table IV.

\section{Conclusion}

\noindent The Schr\"odinger operator $H= -\frac{1}{2}\Delta +V_q(r)$ is of considerable interest in atomic physics.  In this paper we have shown for the two most important cases $q=1$ and $q=2$, that the bound-state eigenequation $H\psi= E\psi$ can be reduced to the problem of solving the confluent Heun differential equation, or the generalized Heun differential equation.  In the present context, the wave function can be written in a form in which analytical solutions with the correct asymptotics involve an unknown factor, which in certain cases can be shown to be a polynomial.  A recently established Asymptotic Iteration Method (AIM) has been  used to determine this factor, and the corresponding eigenvalue, explicitly in cases where the factor is a polynomial, and to construct accurate approximations in other cases.

\section*{Acknowledgments}
\medskip
\noindent Partial financial support of this work under Grant Nos.
GP3438 and GP249507 from the Natural Sciences and Engineering
Research Council of Canada is gratefully acknowledged by two of us
(respectively RLH and NS). KDS acknowledges the Department of
Science and Technology, New Delhi for the award of J.C.Bose
National Fellowship. KDS and NS are grateful for the hospitality
provided by the Department of Mathematics and Statistics of
Concordia University, where part of this work was carried out.
\medskip
\section*{Appendix I: Confluent Heun differential equation}
\noindent In this appendix we summarize some facts about the confluent Heun differential equation, for more
details and further properties, we refer the work of P. P. Fiziev \cite{fiziev} and the references therein.
The confluent Heun's differential equation in simplest uniform shape is written as
\begin{equation}\label{a1}
H''+\left(a+{\beta+1\over z}+{\gamma+1\over z-1}\right)H'+\left({\mu\over z}+{\nu\over z-1}\right)H=0.
\end{equation}
The equation has two regular singularities, at $z = 0$ and $z = 1$, and an irregular singularity
at $z = \infty$.
The constants $\mu$ and $\nu$ are related to the constants $\alpha,\beta,\gamma,\delta,\eta$ in the accepted notation of the confluent Heun function
\begin{equation}\label{a2}
H=He(\alpha,\beta,\gamma,\delta,\eta,z)
\end{equation}
as follows
\begin{equation}\label{a3}
\left\{ \begin{array}{l}
 \delta=\mu+\nu-\alpha\left({\beta+\gamma+2\over 2}\right), \\ \\
 \eta={\alpha(\beta+1)\over 2}-\mu-\left({\beta+\gamma +\beta\gamma\over 2}\right).
       \end{array} \right.
\end{equation}
The confluent Heun functions $He(\alpha,\beta,\gamma,\delta,\eta,z)$ reduces to a polynomial of degree $N\geq 0$ in
the variable $z$ if and only if the following two conditions are satisfied:
\begin{equation}\label{a4}
\left\{ \begin{array}{l}
 {\delta\over \alpha}+{\beta+\gamma\over 2}+N+1=0, \\ \\
 \Delta_{N+1}(\mu)=0
      \end{array} \right.
\end{equation}
where $\Delta_{N+1}(\mu)$ is a three-diagonal determinant given by:
\begin{equation}\label{a5}
\left| \begin{array}{ccccccc}
  \mu-q_1 & 1(1+\beta) &  0&\dots&0&0&0 \\
N\alpha & \mu-q_2+\alpha & 2(2+\beta) &\dots&0&0&0 \\
0     & (N-1)\alpha     & \mu-q_3+2\alpha&\dots&0&0&0\\
\vdots& \vdots&\vdots&\ddots&\vdots&\vdots&\vdots\\
0&0&0&\dots&\mu-q_{N-1}+(N-2)\alpha&(N-1)(N-1+\beta)&0\\
0&0&0&\dots&2\alpha&\mu-q_{N}+(N-1)\alpha&N(N+\beta)\\
0&0&0&\dots&0&1\alpha&\mu-q_{N+1}+N\alpha\\
  \end{array}\right|
\end{equation}
for
$$q_n=(n-1)(n+\beta+\gamma)$$
Under the simultaneous fulfillment of the two additional conditions
(\ref{a4}) the confluent Heun function $He(\alpha,\beta,\gamma,\delta,\eta,z)$ reduces to a polynomial of degree $N$.

\end{document}